# Effect of pH on the complex coacervation and on the formation of layers of sodium alginate and PDADMAC

**Bruno dos Santos de Macedo[1,2]\*, Tamiris de Almeida[3,4]\*, Raphael da Costa Cruz[2], Annibal Duarte Pereira Netto[1], Ladário da Silva[4,5], Jean-François Berret[6] and Letícia Vitorazi[3,4#]**

[1]*Laboratório de Química Analítica Fundamental e Aplicada, Instituto de Química - Universidade Federal Fluminense, R. Outeiro de São João Batista, s/n, Niterói RJ, CEP 24020-141, Brazil.*
[2]*Programa de Pós-Graduação em Química, Instituto de Química - Universidade Federal Fluminense, R. Outeiro de São João Batista, s/n, Niterói RJ, CEP 24020-141, Brazil.*
[3]*Laboratório de Materiais Poliméricos, EEIMVR, Universidade Federal Fluminense, Avenida dos Trabalhadores, 420, Volta Redonda RJ, CEP 27225-125, Brazil.*
[4]*Programa de Pós-Graduação em Engenharia Metalúrgica, EEIMVR, Universidade Federal Fluminense, Avenida dos Trabalhadores, 420, Volta Redonda RJ, CEP 27225-125, Brazil.*
[5]*Laboratório Multiusuários de Caracterização de Materiais, Instituto de Ciências Exatas – Universidade Federal Fluminense, R. Des. Ellis Hermydio Figueira, 783, Volta Redonda RJ, CEP 27213-145, Brazil.*
[6]*Matière et Systèmes Complexes, UMR 7057 CNRS Université Denis Diderot Paris-VII, Bâtiment Condorcet, 10 rue Alice Domon et Léonie Duquet, 75205 Paris, France.*

In this study, we investigated the thermodynamic features of a system based on oppositely charged polyelectrolytes, sodium alginate, and poly(diallyldimethylammonium chloride) (PDADMAC) at different pH values. Additionally, a comparison of the effects of the thermodynamic parameters on the growth of the layers based on the same polymers is presented. For this investigation, different techniques were combined to compare results from the association in solution and co-assembled layers at the silicon surface. Dynamic light scattering (DLS) and isothermal titration calorimetry (ITC) were used for studies in solution, and the layer-by-layer technique was employed for the preparation of the polymer layers. Ellipsometry and atomic force microscopy (AFM) were used to characterize the layer thickness growth as a function of the solution pH, and interferometric confocal microscopy was employed to analyze the topography and roughness of the films. The titration of both polyelectrolytes in two different sequences of additions confirmed the mechanism; it involved a two-step process that was monitored by varying the enthalpy, as determined by ITC experiments, and the structural evolution of the aggregates into coacervates, according to DLS. The primary process is aggregation to form polyelectrolyte complexes having a smaller hydrodynamic diameter, which abruptly transit toward a secondary process because of the formation of coacervate particles that have a larger hydrodynamic diameter. Independent of pH and the sequence of addition, for the first process, both directions are entropically driven. However, the binding enthalpy ($\Delta H_b$) decreased with a decrease in the pH of the solution. The layers grown for the PDADMAC/sodium alginate system demonstrated pH sensitivity with either linear or exponential behavior, depending on the pH values of the polyelectrolyte solutions. The more endothermic process at pH 10 afforded layers with a smaller thickness and with linear growth according to the increase in the number of layers from 5 to 20. Decreases in the pH of the solution resulted in the layers growing exponentially; additionally, a decrease in the $\Delta H_b$ of the association in the solution was observed. The layer thicknesses measured using ellipsometry and AFM data were in good agreement. Additionally, a pH influence on the roughness and topography of the films was observed. Films from basic dipping solutions resulted in surfaces that were more homogeneous with less roughness; in contrast, films with more layers and those formed in a low-pH dipping solution were rougher and less homogeneous.

# To whom all correspondence should be addressed. E-mail: leticiavitorazi@id.uff.br

# INTRODUCTION

Complexes of polyelectrolytes (PECs) are spontaneously formed by attractive interactions when solutions of oppositely charged polymers are mixed.[1] Normally, owing to the equimolarity of charge, soluble PECs produce a phase separation that can be observed by the formation of insoluble aggregates when the equimolarity is achieved, resulting in coacervates or precipitates.[1,2] The nature of the polymers, ionic strength and pH of the solution, can influence the resulting structures, and the mechanisms of the associations between the polymers have been studied by relating the thermodynamic parameters with the structural changes.[2-4] Polyelectrolyte multilayers (PEMs) can be obtained by the layer-by-layer technique, which is based on the physical adsorption of oppositely charged species via electrostatic interactions.[5] This technique has been widely used owing to its low cost, large-scale reproducibility, simple manipulation, and ease in making films with a controlled thickness, composition, and structure.[6]

Applications of PEMs extend their scope owing to the possibility of controlling assemblies of polymers to produce different materials. They are useful as agents in the modification of surfaces[7], can be used as platforms for the delivery of nucleic acids[8], for substrate-mediated delivery in biomedical applications[9], in the preparation of biocompatible bacterial-resistant films[10,11], and can be used as materials for tissue engineering.[12,13] Previously, polyelectrolyte multilayers (PEMs) have been formed by the alternative deposition of polyelectrolytes[14] on two-dimensional (2D) or three-dimensional (3D) surfaces.[13] Recently, polyelectrolyte-coated microcapsules for the encapsulation of a self-healing agent for the corrosion protection of steel[15] and capsules with antibacterial activity have been proposed.[16] Laugel and co-workers[14] compared several systems to predict the film growth regime by the thermodynamic parameters of polyelectrolyte complexation. In their study, they associated an endothermic or weak exothermic process with exponentially growing films and a strong exothermic process with linear film growth. Additionally, the exponential growth of layers was associated with polymers that interact weakly, resulting in less structured films. When the growth regime becomes exponential, the film formation process is dynamic, and there may be diffusion between layers.[14,17,18] In this regime, which is common for polypeptides and polysaccharides, the film is less structured and more hydrated, and its thickness can reach micrometers after deposition of only a few tens of layers.[19] However, for the linear growth regime, a more stratified structure has been observed in which the layers are finely accommodated.[14] For films obtained using poly(diallyldimethylammonium chloride), PDADMAC and poly(styrenesulfonate) (PSS) polymers, both growth regimes have been observed according to the salt concentration and presence of a buffer.[20,21]

In this study, we evaluate the mechanism of the association of the polyelectrolyte, PDADMAC, with sodium alginate. Sodium alginate is a biodegradable polymer that exhibits liquid–gel behavior in aqueous solutions and has been studied for several years in the biomedical field, especially in drug delivery and tissue engineering, where it can be used in the regeneration of skin tissue, muscles, and nerves. Sodium alginate has also been used in the development of biosensors from biopolymer composites, biodegradable textile yarns, and in enzyme mobilization.[22] We investigated the mechanism of association of these polyelectrolytes in solution according to the thermodynamic parameters and structural modification of the system by following a methodology of titration that has been previously described.[2,23,24] These studies were performed to evaluate the influence of the pH on the PEM formation with prior understanding of the co-assembly of the polymeric pairs, weak versus strong, in solution.

## MATERIALS AND METHODS

### Materials

A polymer solution of poly(diallyldimethylammonium chloride) at 35% (weight/volume) in water (PDADMAC, $M_w < 100\ 000$ g mol$^{-1}$) and copolymer sodium alginate in powder form ($M_v \approx 51\ 000$ g mol$^{-1}$; see Supporting Information S1) were both purchased from Sigma-Aldrich (USA) and used as received. The pH of the polyelectrolyte solutions or water to remove the free polymer chains from the wafer silicon surface, was adjusted by adding few drops of diluted solutions at different concentrations (0.005, 0.01, 0.05, 0.1, and 0.25 mol L$^{-1}$) of hydrochloric acid (HCl, analytical grade reagent, 36.5 %w/w, Proquimios, Brazil) or sodium hydroxide (NaOH, analytical grade reagent, 97%, Sigma-Aldrich, USA) that were previously prepared in ultrapure water. The cleaning of the silicon wafers was performed using solutions (3:7, v/v) of hydrogen peroxide (H$_2$O$_2$, 30 volumes, analytical grade reagent, Dinâmica, Brazil) and ammonium hydroxide (NH$_4$OH, analytical reagent, 24% w/w, IMPEX, Brazil) or sulfuric acid (analytical grade reagent, 18% w/w, SCIAVICCO, Brazil). All aqueous solutions were prepared in ultrapure water with a resistivity of 18.2 MΩ cm at 25 °C, obtained using a water purification system (Arium Confort II, Sartorius, Brazil). The polymer structures are shown in Figure 1A. Figure 1B shows the two types of titration experiments. They are denoted as $Z_{(+/-)}$ = [PDADMAC]/[sodium alginate] or $Z_{(-/+)}$ = [sodium alginate]/[PDADMAC]. In all experiments, the charge ratio, $Z_{(+/-)}$ or $Z_{(-/+)}$, was assumed to be the ratio of the molar concentrations of PDADMAC to sodium alginate, according to the molar masses of the polymer unit (198.11 g mol$^{-1}$ and 161.67 g mol$^{-1}$), for sodium alginate and PDADMAC, respectively. For example, a solution of 0.020 mol L$^{-1}$ of PDADMAC contains 3.2 g L$^{-1}$. The degree of protonation of sodium alginate was considered as 100% for the prepared solution and obtained results. The subscripted notation (+/-) indicates the titration of sodium alginate with the PDADMAC solution for Type I experiment, and (-/+) indicates the titration of PDADMAC solution using the sodium alginate solution for Type II experiment, as shown in Figure 1B.

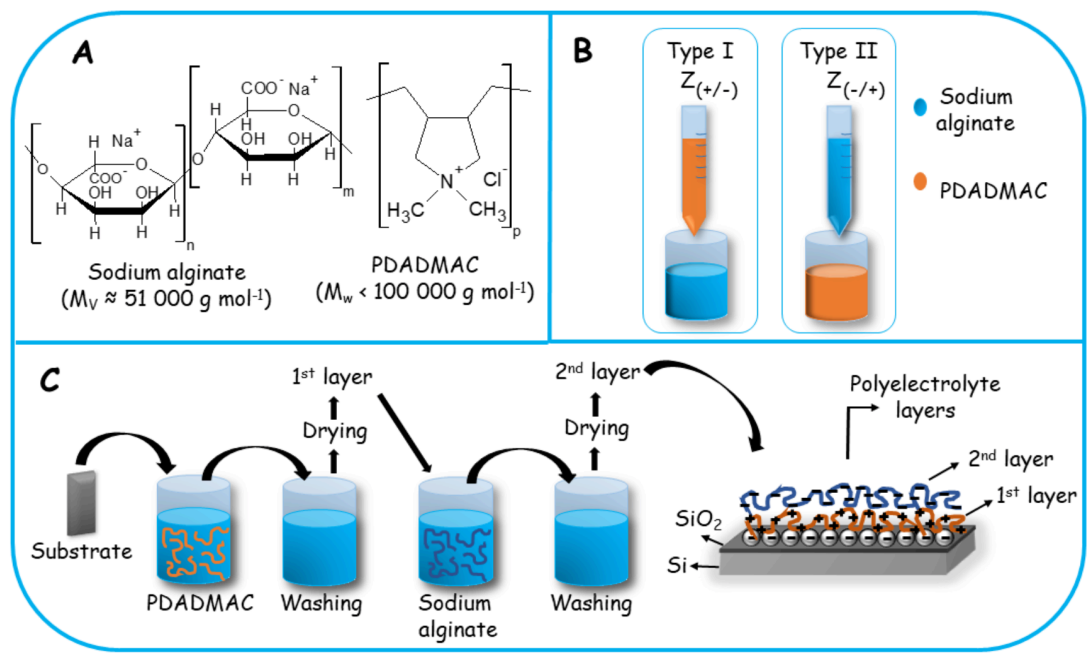

***Figure 1.*** *(A) Structure of the sodium alginate and PDADMAC polymers, (B) Type I and II titration experiments, and (C) schematic of the two-layers film preparation applied via the layer-by-layer technique using a silicon substrate and alternating dipping solution of PDADMAC or sodium alginate.*

## Titration protocols

The experiments were performed using stock polymer solutions at same pH values before the mixture for all studies that employed titration protocols. The concentrations of the titrant solutions were always ten times higher than that of the titrated solutions and injected stepwise into a solution containing the oppositely charged polymer. Thus, in all cases, PDADMAC or sodium alginate solutions (0.0200 mol L$^{-1}$) were added stepwise to sodium alginate or PDADMAC solutions (0.00200 mol L$^{-1}$), respectively (Figure 1B). Microcalorimetry, pH, dynamic light scattering and electrophoretic mobility measurements were performed during the titrations.

## Isothermal titration calorimetry

Isothermal titration calorimetry was performed using a calorimeter (Microcal® VP-ITC, Northampton, MA) with a 1.464-mL cell working at 25° C and an agitation speed of 210 rpm. The syringe and the measuring cell were filled with degassed solutions of the polyelectrolytes PDADMAC or sodium alginate at pH values of 4, 5, 7, and 10. All experiments consisted of a preliminary injection of 2 μL (removed for data treatment), followed by 29 injections of 10 μL with time intervals of 350 s between each injection. Experiments were performed in triplicate but a single curve was used to represent the results, and average data were used to compare the results.

## Isothermal titration calorimetry fundamentals and data treatment

For macromolecule M and ligand X to form a complex MX, it is possible to write an equation for the chemical equilibrium, where $K_b$ is the binding constant:

$$M + X \rightleftharpoons MX \qquad (1)$$

When a macromolecule of initial concentration, Mt, is titrated with a ligand of constant concentration Xt to form complex MX, the total heat (kJ mol$^{-1}$) released or absorbed per injection of titrant can be estimated from the multiple non-interacting binding sites (MNIS) model. In this model, the total heat is proportional to the stoichiometry of the reaction denoted by $n_b$, as proposed by Wiseman.[25] The concentrations of free macromolecule [M], free ligand [X] and complex formed [MX] are variables in the system, which we can rewrite using the mass balances (Eq. 2 and 3) and the binding constant $K_b$ as follows:

$$n_b Mt = [M] + [MX] \qquad (2)$$

$$Xt = [X] + [MX] \qquad (3)$$

$$K_b = \frac{[MX]}{[M][X]} \qquad (4)$$

By rewriting these equations to isolate variable [MX], Eq. 5 is obtained:

$$[MX] = \frac{n_b Mt + Xt + 1/K_b - \sqrt{(n_b Mt + Xt + 1/K_b)^2 - 4 n_b Mt Xt}}{2} \qquad (5)$$

In ITC experiments, the titration progress can be monitored as a function of the molar ratio, which is defined as $Xr = Xt / n_b Mt$. If we multiply some terms from Eq. 5 for $n_b Mt / n_b Mt$, it is possible to obtain:

$$\frac{[MX]}{n_b Mt} = \frac{1 + Xr + r - \sqrt{(1 + Xr + r)^2 - 4Xr}}{2}, \qquad (6)$$

$$\text{where } r = \frac{1}{n_b K_b Mt} \qquad (7)$$

The total heat released or absorbed per mole of titrant $n_{Xt}$ can be written as a function of the molar ratio by Eq. 8. By substituting Eq. 6 into Eq. 8 and performing the derivative, it is possible to obtain the general equation (Eq. 9), which represents the total heat released or absorbed per injection of titrant, in kJ mol$^{-1}$.

$$\frac{dq_T}{dn_{Xt}} = \Delta H_b \frac{d}{dXr} \frac{[MX]}{n_b Mt} \qquad (8)$$

$$\frac{dq_T}{dn_{Xt}}(X_r, n_b, K_b) = \frac{1}{2} \Delta H_b \left( 1 + \frac{1 - Xr - r}{\sqrt{(1 + Xr + r)^2 - 4Xr}} \right) \qquad (9)$$

By plotting the experimental data and using the molar ratio to determine the best set of parameters that fit the experimental data to the theoretical curve obtained, a thermogram is obtained from Eq. 9. Generally, an iterative method such as a least-squares optimization technique is used to determine the parameters of $n_b$, $r$, and binding enthalpy $\Delta H_b$. In this study, we employed a modified version of the MNIS model[25] to treat the experimental data as previously proposed[2], which will be called the MMNIS model. Equations 10 and 11 are based on Eq. 9 and provide the total heat per injection in the first step and second step, respectively:

$$\frac{dq_T^A}{dn_{Xt}}(Xr, n_b^A, K_b^A) = \frac{1}{2}\Delta H_b^A\left(1 + \frac{1 - Xr - r^A}{\sqrt{(1 + Xr + r^A)^2 - 4Xr}}\right) \quad (10)$$

$$\frac{dq_T^B}{dn_{Xt}}(Xr, n_b^B, K_b^B) = \frac{1}{2}\Delta H_b^B\left(1 + \frac{1 - Xr - r^B}{\sqrt{(1 + Xr + r^B)^2 - 4Xr}}\right) \quad (11)$$

where the derivatives: $\frac{dq_T^A}{dn_{Xt}}(Xr, n_b^A, K_b^A)$ and $\frac{dq_T^B}{dn_{Xt}}(Xr, n_b^B, K_b^B)$ represent the total heat per injection in the first step (aggregation to form the polyelectrolyte complexes) and second step (formation of coacervates), respectively. Furthermore, $r^A = 1/n_b K_b^A Mt$ and $r^B = 1/n_b K_b^B Mt$ are parameters directly related to the binding constants.[26] $K_b^A$ is the binding constant in the first step, and $K_b^B$ is the binding constant in the second step.

The first-step process is represented by a curve with a decreasing sigmoidal function, and the second-step process is represented by a curve with a Gaussian shape. Here, we are assuming that the first- and second-step processes are sequential, i.e., the contribution of the second-step process to the total heat becomes significant only when the contribution of the first-step process decreases.[2] Thus, the total released or absorbed heat per injection during titration can be written as a combination of the heat of the first-step process (Eq. 10) plus the heat of the second-step process (Eq. 11) times multiplied by parameter $\alpha(Xr)$, giving Eq. 12:

$$\frac{dq_T}{dn_{Xt}}(Xr) = \frac{dq_T^A}{dn_{Xt}}(Xr, n_b^A, K_b^A) + \alpha(Xr)\frac{dq_T^B}{dn_{Xt}}(Xr, n_b^B, K_b^B) \quad (12)$$

where $\alpha(Xr)$ is a step function applied in the second term to control heat contributions and make it sequential, thus relating the calorimetric behavior of each step to structural changes.[3] A step function can assume several forms, but in this study, the step function has the form:

$$\alpha(Xr) = \frac{1}{1 + \exp[-(X_r - n_b^A)/\sigma]} \quad (13)$$

where $n_b^A$ is the stoichiometry of the first step and $\sigma$ represents a constant that is related to the laterality of the function.

**Dynamic light scattering and electrophoretic mobility**

Dynamic light scattering (DLS) and electrophoretic mobility measurements were performed using a zetasizer (Malvern Instruments®, Nano-ZS, United Kingdom) at 25° C. Solutions (7 mL) of sodium alginate (0.00200 mol L$^{-1}$) with pH adjusted to 4 or 10 were titrated with small aliquots (50 μL) of PDADMAC (0.0200 mol L$^{-1}$) at the same pH under continuous stirring. The inverse experiments, when PDADMAC solutions were titrated with sodium alginate solutions, were also conducted in the same manner. Light scattering data were expressed as the Z-average (D$_H$, nm), and the scattered intensity (Int, kc s⁻¹), while electrophoretic mobility measurements were expressed as the zeta potential (ZP, mV).

## Layer-by-layer and film characterization
### Cleaning of the silicon substrate and deposition of layers

Silicon substrates (~1 × 1.5 × 0.53 cm) were cleaned carefully by immersion in solutions of sulfuric acid or ammonium hydroxide containing hydrogen peroxide, according to a previous method[27], as detailed in Supporting Information S2. The multilayers were prepared manually by immersing the previously cleaned silicon substrate in a polyelectrolyte solution (layer-by-layer method). For the growth of the first layer, the cleaned silicon substrate was immersed in a PDADMAC solution (0.0200 mol L⁻¹) for 10 min. After this period, it was washed (by immersion) for 30 s in a dilute NaOH aqueous solution at pH 10 and dried under a gentle $N_2$ flow. Subsequently, for the growth of the second layer, the silicon substrate was immersed in a sodium alginate solution (0.0200 mol L⁻¹) for 10 min, washed by immersion in water at pH 10 for 30 s, and dried under a gentle $N_2$ flow. The layer deposition was repeated by immersing 1–5, 10, 15, and 20 times, and each immersion was considered to deposit a layer (Figure 1C). The layer deposition was also conducted at pH 3 and 6 using the same 0.0200 mol L⁻¹ solutions followed by washing with water at the same pHs. All layer experiment steps were performed at room temperature around 24-25 °C.

## Ellipsometry

The film thickness was measured in an ellipsometer (model SOPRA GES 5S, SEMILAB®, Hungary), which used a Xenon lamp with a spectral range of 200–1000 nm. Measurement of ellipsometric parameters Ψ and Δ was performed by applying an incidence angle of 70° with a microspot.[28-30] The ambient temperature was 23 °C during all measurements. Because ellipsometry is an indirect technique, computational modeling of the data was necessary to obtain the optical parameters and film thickness.[31] Spectroscopy ellipsometry analyzer (SEA) software from SEMILAB was used. The data modeling considered the substrate, which was formed by silicon and silicon oxide as the basis for the film growth, and the subsequent polyelectrolyte layers. The cleaned silicon substrate was previously characterized before starting the growth of the polyelectrolyte layers.[32] Measurement and modeling were performed for each polymer layer that deposited at the silicon substrate. To verify the quality of curve fitting of ellipsometric parameters Ψ and Δ as a function of λ, statistical parameters $R^2$ (coefficient of determination) or RMSE (root mean square error) were evaluated.[33] For measurements of 1 to 5 layers at pH 10, modeling of the layers was performed using the Cauchy, Lorentz, Gauss, and Cauchy with Urbach tail dispersion laws, and the refractive index was fixed closer to 1.46 at 632.8 nm wavelength, as described in the literature[27,34] to obtain the polymer layer thickness. For the other layers, modeling was performed using Cauchy, Tauc–Lorentz, Gauss, and Cauchy with Urbach tail and EMA dispersion laws, and the refractive index was not fixed.

**Atomic force microscopy**

An atomic force microscope (model FlexAFM V2, Nanosurf ®, USA) was used to analyze the topography and study the thickness of the films and to compare with ellipsometric data. The measurements were performed using a Tap190A1-G cantilever in contact mode. The thickness determination was obtained by the scratching technique, and the measurements were made by scanning in and out of the scratch. All measurements were performed in air under ambient conditions, and each image scanning on a sample corresponded to 256 x 256 pixels. The scanning speed used was 0.7 s per line, and the size of the scanned area was $30 \times 30$ μm. The measurements were made in four different scratching areas, and the thickness was calculated from a mean of 15 spots taken in each area.

**Interferometric confocal microscopy**

An interferometric confocal microscope (Leica Microsystem, DCM 3D, Germany) was used to analyze the topography and roughness of the films. A ten times magnification lens and blue LED at 460 nm were employed for the measurements. Roughness measurements were based on DIN 4768[35], and the data were analyzed according to a completely randomized design (DIC). An analysis of variance (ANOVA) and subsequently an average test (Tukey) were performed at a 5% level of significance. The program used for statistical analyses was SISVAR®

# RESULTS AND DISCUSSION

**Titration curve analysis based on isothermal titration calorimetry**

Figures 2a and b show the binding isotherms obtained for the studied titrations of sodium alginate and PDADMAC. A decrease in the heat release occurred with increasing of $Z_{(+/-)}$ or $Z_{(-/+)}$ ratios for Type I and Type II experiments, respectively. Initially, the enthalpy variation indicated that both experiments were endothermic. Furthermore, we observed the influence of pH on the enthalpy curves. Although the same concentrations were employed for each set of experiments, there were no superimpositions of curves in both sets of experiments, showing the effect of pH on the obtained results. The displacements of the enthalpy curve to the lowest values of $Z_{(+/-)}$ with the pH decrease and change in the enthalpy values may be associated with the pKa of the sodium alginate polymer owing to the degree of protonation that influences the electrostatic complexation.

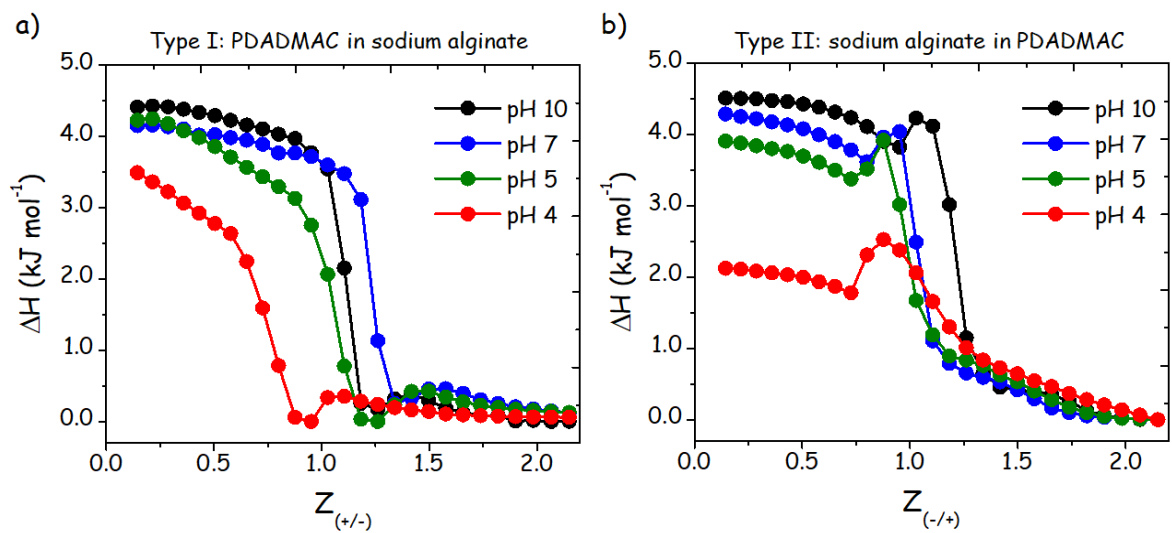

***Figure 2.*** *ITC curves obtained at 25 °C (a) for the titration of sodium alginate using PDADMAC (Type I experiments) and (b) for the titration of PDADMAC in sodium alginate (Type II experiments) at different pH values.*

No measurable variation in pH was observed when the titration was performed at pH 10. This occurred because, at pH 10, the concentration of $H_3O^+$ was not significantly high, which indicated that all alginate sites were deprotonated. In contrast, at pH values below 7, the concentration of $H_3O^+$ increased, reducing the deprotonated alginate sites and causing a measurable pH variation along the titration.[25] For the pH values of 7, 5, and 4, the titration progress caused a decrease in the pH, which, later, stabilized below the initial value. However, when PDADMAC was titrated with alginate, the pH decreased along the titration with final stabilization near the initial values. The pH behavior along the titrations is detailed in Supporting Information S3. Nevertheless, Type I curves show a second process that includes a peak due to an exothermic process after the titration had been completed at $Z_{(+/-)}$ ratios between 0.9 and 1.4 (Figure 2a), and this process was also observed in Type II experiments at $Z_{(-/+)}$ ratios between 0.9 and 1.2 (Figure 2b). However, in this case, the peak represents an endothermic process.

In Type I experiments, the exothermic peak for the experiment at pH 4 was present at the lowest values of $Z_{(+/-)}$, (~1). For the other curves obtained at pH 5, 7, and 10, the exothermic peak occurred at the highest $Z_{(+/-)}$ ratios. A similar behavior was observed for Type II experiment: there was a displacement in the endothermic peak for higher $Z_{(-/+)}$ ratios with an increase of the pH of the solution. The decrease in the endothermic peak appeared at $Z_{(-/+)} \sim 1.0$ for pH 10, $Z_{(-/+)} \sim$ 0.90 for pH 7, and 0.88 for pH 5 or 4. For both Type I and Type II results, two contributions to the total enthalpy included two processes occurring during titration that evolved as an exothermic peak or endothermic peak. These results are similar to those obtained previously for the complexation of poly(sodium acrylate) (PANa) with PDADMAC.[2] Liu and co-workers[36] associated the endo- or exo-thermic events as the magnitude of the second peak with the energy cost to reach its condensation in neutral coacervated droplets, according to the charge of the aggregates previously formed.

**Relationship between thermodynamic titration and structures of complexes in solution**

DLS measurements were performed under the same conditions as the ITC experiments, to compare the structure of the species evolved during the titration with the thermodynamic data from the complexation. Figures 3a and 3b show the combined data from the ITC and DLS measurements for experiments conducted at pH 10. These figures display the binding enthalpy, $\Delta H$ (kJ mol$^{-1}$), the intensity of light scattering, Int (kc s$^{-1}$), the hydrodynamic diameter, $D_H$, (nm), and the zeta potential, ZP (mV), as a function of the mixing ratio at pH 10 for Type I and Type II experiments.

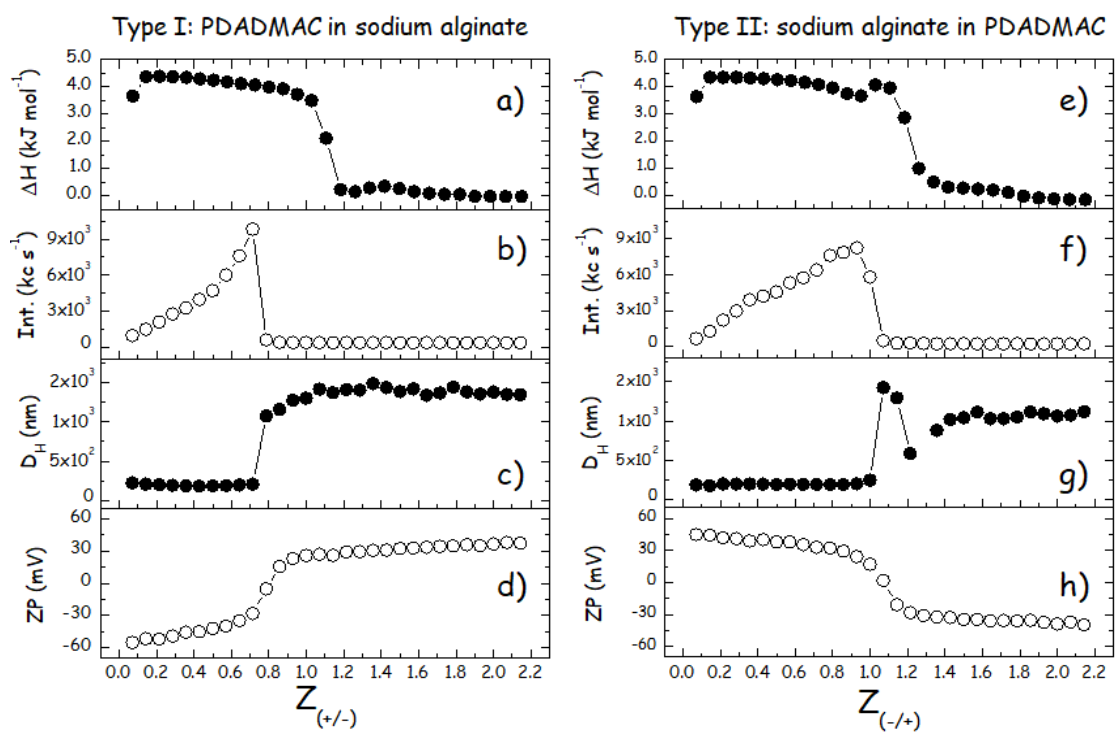

***Figure 3.*** *ITC curves, the intensity of light scattering, hydrodynamic diameters and zeta potentials at 25 °C (a–d) for the titration of sodium alginate using PDADMAC for Type I experiments and (e–h) for the titration of PDADMAC using sodium alginate for Type II experiments at pH 10.*

The comparison of Figures 3b and 3f for Type I and II experiments shows that the scattered intensity increased progressively to $10^4$ kc s$^{-1}$ up to $Z_{(+/-)} = 0.7$ and up to $Z_{(-/+)} = 0.9$, respectively. Moreover, the hydrodynamic diameter was approximately constant ($D_H \sim 200$ nm) in both Figures 3c and 3g until $Z_{(+/-)}$ and $Z_{(-/+)}$ showed similar values. Titration achieved an isoelectric point at $Z_{(+/-)} = 0.8$, as shown in Figure 3d. Before this point, the surface charge of the aggregates was negative owing to an excess of the polyanion sodium alginate, and afterwards, the coacervates had a positive charge because of the excess of polycation PDADMAC. For both experiments types, a clear separation of the process occurred for the system that was represented by the specific value of Z at the critical molar ratio, which was near the isoelectric point of the titration.

These phenomena indicate a transition of PDADMAC/sodium alginate aggregation to form co-acervates, like those previously obtained for the titration of polyacrylic acid and PDADMAC.[2,36] In Type I and II experiments, for $Z_{+/-}$ or $Z_{-/+}$ ratios that were larger than the critical molar ratio, the diameters of the particles increased rapidly, followed by a decrease in the scattered intensity up to a plateau. The hydrodynamic diameter was approximately constant ($D_H \sim 200$ nm) for Type I and II titrations before the critical molar ratio. Subsequently, it increased rapidly, reaching values of $D_H$ that were larger than 1500 nm. However, above this, DLS was not appropriate to evaluate exactly the sizes of the structures formed, and our data indicated the formation of large structures.[2,36]

Zeta potential measurements obtained for Type I titration indicated that the aggregates were initially negatively charged (−60 mV), but with the evolution of the titration, it increased to ca. 0 mV, indicating an isoelectric point. For Type II experiment, a similar behavior was observed, but with the opposite charge, aggregates formed were positively charged (+45 mV), and the zeta potential decreased to 0 mV at the isoelectric point. Thus, the mixing sequence influenced the obtained structures. The total process was endothermic for both titration modes, and, in general, they exhibited similar features; however, the sizes, scattered light intensities, and hydrodynamic diameters were different. Figures 4a–d and 4e–h show the data obtained from the ITC and DLS measurements for experiments conducted at pH 4. These figures display the binding enthalpy, ΔH, the scattering intensity, Int, the hydrodynamic diameter, $D_H$, and the zeta potential, ZP, as a function of the mixing ratio at pH 4.

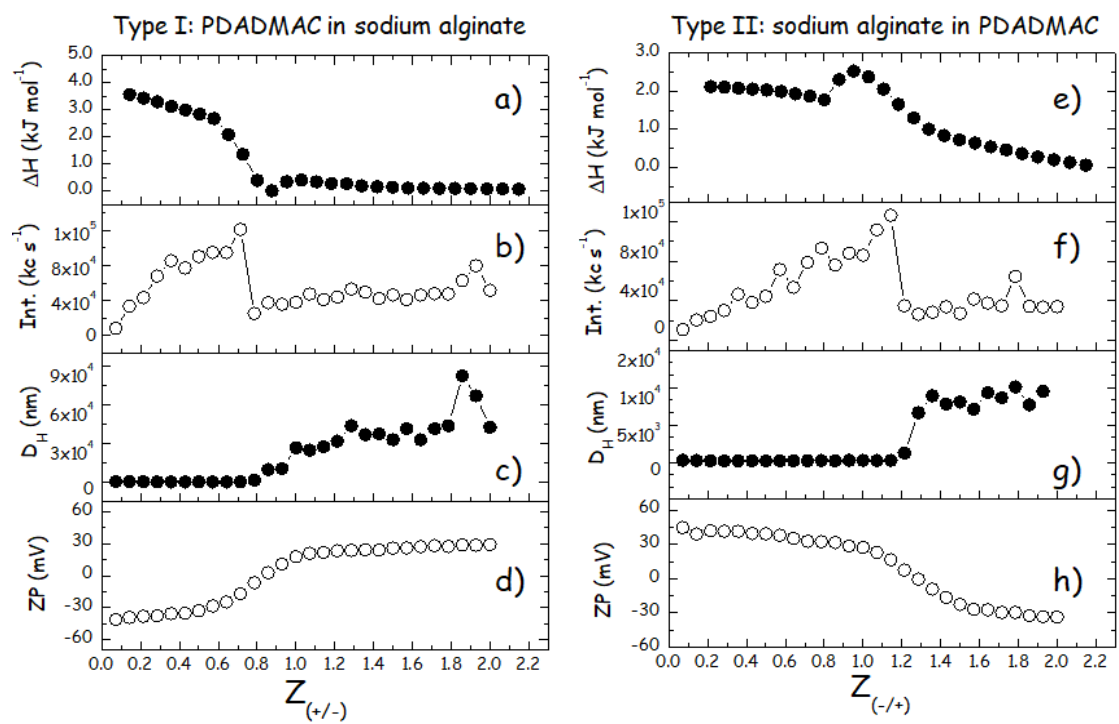

**Figure 4.** ITC curves, light scattering intensities, hydrodynamic diameters, and zeta potentials at 25 °C (a–d) for the titration of sodium alginate using PDADMAC in Type I experiment and (e–h) for the titration of PDADMAC using sodium in Type II experiment at pH 4.

The comparison of Figures 3 and 4 showed that the curves behave differently according to the pH condition. However, the enthalpy kept exhibiting an exothermic peak during Type I experiment and an endothermic peak during Type II experiment, indicating that they have the same signal with pH change. For both processes, the variation in the enthalpy during the first process was positive, indicating that the total process was endothermic. The scattered light intensity exhibits a plateau at approximately $4 \times 10^4$ kc s$^{-1}$ above $Z_{(+/-)} = 0.8$ for Type I and above $Z_{(-/+)} = 1.2$ for Type II experiment, indicating the formation of large species that absorbed light or partially decanted from the solution before the critical molar ratios had occurred, followed by a progressive increase in the scattered light intensity up to $1.2 \times 10^5$ kc s$^{-1}$ in both experiments.

For Type I and II experiments, the hydrodynamic diameters were approximately constant at $D_H \sim$ 200 nm before the critical molar ratio. Subsequently, the size increased significantly, similarly to that observed in Figure 3. $D_H$ increased rapidly until reaching $10^4$ nm at $Z_{(+/-)} = 0.8$ for Type I and $Z_{(-/+)} = 1.2$ for Type II experiments. Zeta potential measurements for Type I titration indicated that aggregates were initially negatively charged with −35 mV because of an excess of sodium alginate at the beginning of the titration. For Type II experiment, the aggregates were initially positively charged with 50 mV because of an excess of PDADMAC at the beginning of the titration. Light scattering and zeta potential experiments were performed only at pH 4 and 10 to determine the structural changes in the systems that presented more variation in the enthalpy curves.

**Thermodynamic parameter analysis based on ITC data**

Figures 5 and 6 display the binding isotherms at different pH values for Type I and II experiments, respectively. Blue points represent the experimental data from ITC. Additionally, the total heat from the MMNIS model is shown by a red line. This function has two components (gray lines), which are further discussed below. The step function is not shown in these figures.

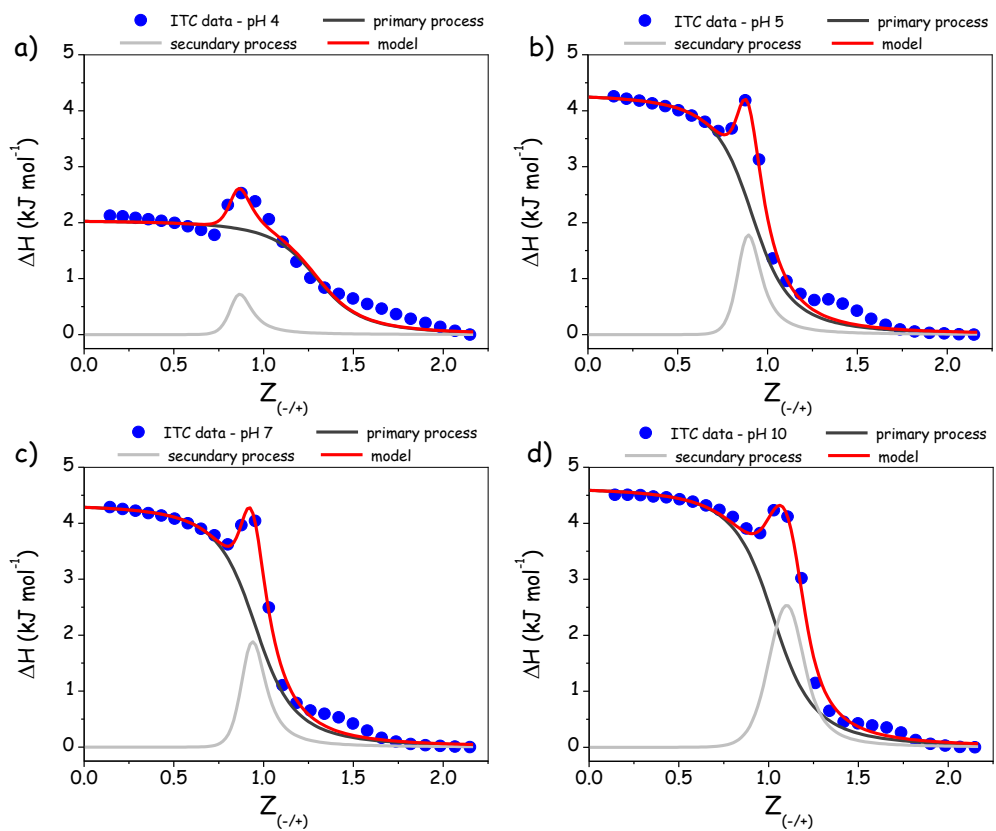

***Figure 5***. *Binding isotherms for sodium alginate in PDADMAC titrations at pH (a) 4, (b) 5, (c) 7, and (d) 10 for Type II experiments at 25 °C. The uncertainty in the determination of the fitting parameters is approximately 5%.*

Upon comparing Figures 3 and 4 with Figures 5 and 6, the structural modifications of the system that evolve as two-step processes were found to contribute to the total enthalpy of the association of the polyelectrolytes. From the binding isotherms, we observed that each process has its thermodynamic parameters that can be treated mathematically, considering that the total heat exchange is the sum of two contributions.[2] Visual inspection of Figure 5 shows slight differences between the data and the model in the range of Z = 1.2–1.8, with stronger effects at pH 5 and 7. This deviation could indicate that a third process occurred during the titration. Of note, Type II thermograms are well fitted by the MMNIS model (Eq. 12) and do not display this anomaly.

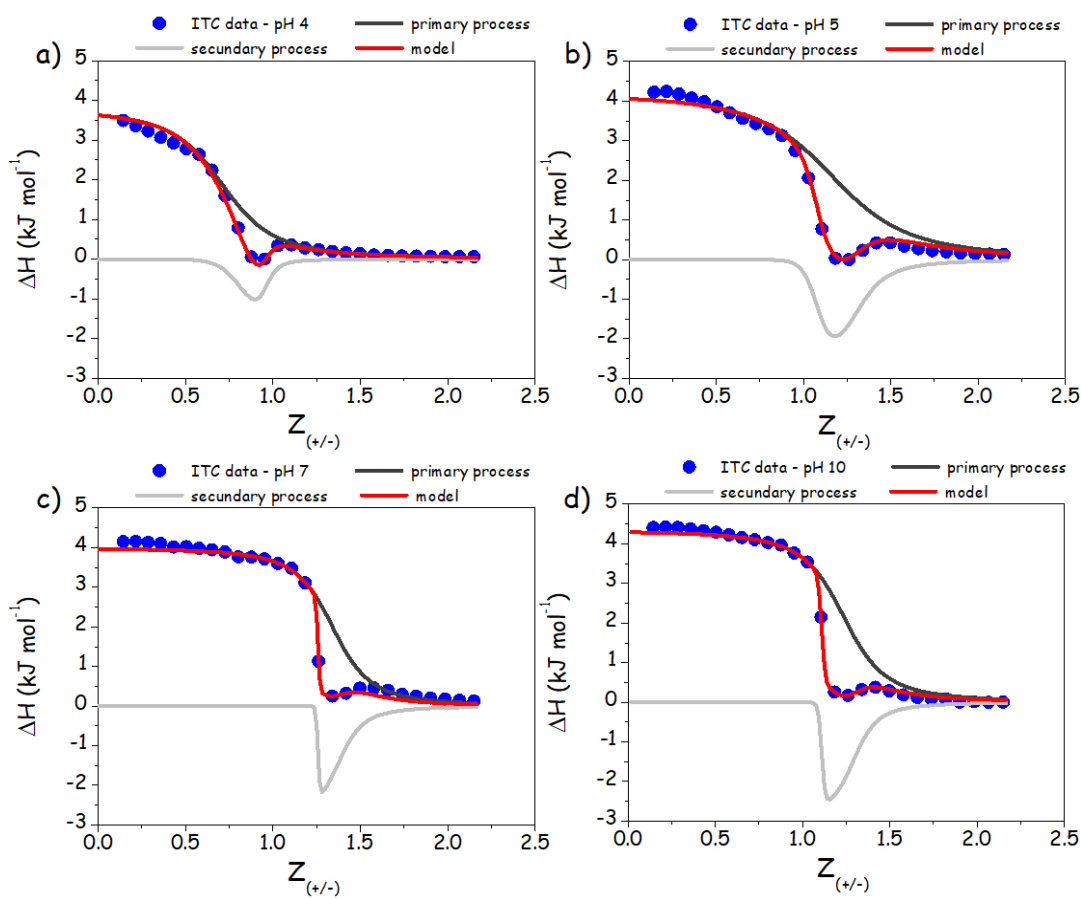

***Figure 6***. *Binding isotherms for PDADMAC in sodium alginate titrations at pH (a) 4, (b) 5, (c) 7, and (d) 10 for Type I experiments at 25 °C. The uncertainty in the determination of the fitting parameters is approximately 5%.*

The total heat exchange from the MMNIS model is represented by the red curve in Figures 5 and 6, which is assumed to occur in two sequential events. The concept of sequential processes applied in this context is not new. Other researchers have used this approach to study different systems, such as to obtain the thermodynamic parameters from polyethyleneimine–DNA binding and DNA condensation[37], in the study of poly-micelle formation[38], and for the complexation of PANa with PDADMAC.[2] Even though several systems have been studied, the effect of pH on the PDADMAC and sodium alginate complexation has not yet been reported in the literature, and the expected behavior is not obvious. For example, the complexation between PANa and PDADMAC at 0.5 mol L⁻¹ of NaCl as a function of pH was entropically driven in basic solution at pH 10 and enthalpically driven toward acidic pH conditions (pH 3 and 6).[39] This thermodynamic behavior was not observed for the sodium alginate and PDADMAC system, as shown in Figures 5 and 6, where the enthalpy was positive in all cases. Thus, all processes in this study were entropically favorable, in contrast to that reported by Alonso and co-workers[39] for PANa and PDADMAC.

Table 1 shows the set of parameters obtained for sodium alginate and PDADMAC complexation from Type I and Type II experiments using Equations 11–13. Equations 14 and 15 allow the calculation of the variation in entropy and free Gibbs energy as they relate to the binding process.

$$\Delta G_b = - RT\ln K_b \qquad (14)$$

$$\Delta S_b = \frac{\Delta H_b - \Delta G_b}{T} \qquad (15)$$

Using these two equations, we could obtain data shown in the fifth and sixth columns of Table 1.

**Table 1.** Thermodynamics parameters for the first-step (A) and second-step (B) processes obtained from adjustment of the ITC curves by Equation 12. $\Delta H_b$, $K_b$, $n_b$, $\Delta G_b$, and $\Delta S_b$ denote the binding enthalpy, binding constant, stoichiometry, Gibbs free energy, and entropy changes, respectively, at 25 °C. The uncertainty in the determination of the fitting parameters is approximately 5%.

| First process (A) | $\Delta H_b^A$ | $K_b^A$ | $n_b^A$ | $\Delta G_b^A$ | $\Delta S_b^A$ |
|---|---|---|---|---|---|
| **Type II** Alginate in PDADMAC | (kJ mol$^{-1}$) | (M$^{-1}$) | | (kJ mol$^{-1}$) | (J mol$^{-1}$ K$^{-1}$) |
| pH 4 | 2.1 | $2.06 \times 10^4$ | 1.29 | −24.6 | 89.5 |
| pH 5 | 3.9 | $3.62 \times 10^4$ | 1.01 | −26.0 | 100.0 |
| pH 7 | 4.4 | $3.79 \times 10^4$ | 0.97 | −26.2 | 102.0 |
| pH 10 | 4.7 | $3.52 \times 10^4$ | 1.04 | −26.0 | 103.0 |
| **Type I** PDADMAC in alginate | | | | | |
| pH 4 | 3.8 | $1.17 \times 10^4$ | 0.74 | −23.2 | 90.6 |
| pH 5 | 4.0 | $9.38 \times 10^4$ | 1.03 | −28.3 | 109.3 |
| pH 7 | 4.1 | $9.38 \times 10^4$ | 1.08 | −28.4 | 109.4 |
| pH 10 | 4.3 | $3.14 \times 10^4$ | 1.25 | −25.7 | 110.0 |
| **Second process (B)** | $\Delta H_b^B$ | $K_b^B$ | $n_b^B$ | $\Delta G_b^B$ | $\Delta S_b^B$ |
| **Type II** Alginate in PDADMAC | (kJ mol$^{-1}$) | (M$^{-1}$) | | (kJ mol$^{-1}$) | (J mol$^{-1}$ K$^{-1}$) |
| pH 4 | 1.8 | $6.55 \times 10^4$ | 0.89 | −27.5 | 98.1 |
| pH 5 | 4.1 | $2.04 \times 10^5$ | 0.86 | −30.3 | 116.0 |
| pH 7 | 4.2 | $1.81 \times 10^5$ | 0.96 | −30.0 | 115.0 |
| pH 10 | 4.3 | $1.49 \times 10^5$ | 1.17 | −29.5 | 113.0 |
| **Type I** PDADMAC in alginate | | | | | |
| pH 4 | −1.7 | $2.27 \times 10^4$ | 0.96 | −30.6 | 97.0 |
| pH 5 | −2.8 | $3.22 \times 10^4$ | 1.31 | −25.7 | 77.0 |
| pH 7 | −2.9 | $8.76 \times 10^4$ | 1.38 | −28.2 | 85.0 |
| pH 10 | −3.0 | $9.14 \times 10^4$ | 1.29 | −28.3 | 85.1 |

As shown in Table I, there were some changes in the parameters according to the mixing protocol. This could be due to the different affinity for the binding sites depending on the pH and the excess of PDADMAC or sodium alginate in the medium, which directly influenced the enthalpy and entropic terms. These quantitative results confirmed the previous qualitative evaluations. The first-step process of the complexation between sodium alginate and PDADMAC was endothermic for both mixing sequences and in the entire range of investigated pH values. The ranges of heat change ($\Delta H_b^A$) were $4.65 \pm 0.20$ to $2.07 \pm 0.12$ kJ mol$^{-1}$ for Type II and $4.34 \pm 0.15$ to $3.79 \pm 0.18$ kJ mol$^{-1}$ for Type I experiments. Furthermore, the heat absorbed decreased with decreasing pH.

While the first-step process showed apparent single behavior, the second step included two steps, depending on the addition sequence during titration. For Type II experiments, the coacervation was endothermic, with $\Delta H_b^B$ varying between $4.27 \pm 0.18$ and $1.76 \pm 0.11$ kJ mol$^{-1}$, while for Type I, the process was exothermic with $\Delta H_b^B$ varying between $-1.65 \pm 0.10$ and $-2.96 \pm 0.14$ kJ mol$^{-1}$. The uncertainty associated with the determination of the fitting parameters fluctuated around 5% in agreement with the slight increase with pH. An evaluation of the influence of the enthalpic and entropic terms upon varying the Gibbs free energy was obtained using Equations 14 and 15. The variation in the Gibbs free energy was negative for both types of experiments, indicating favorable processes. For Type II experiments, $\Delta G_b^A$ varied between $-26.0$ and $-24.6$ kJ mol$^{-1}$, and for Type I experiments, $\Delta G_b^A$ varied between $-26.2$ and $-21.3$ kJ mol$^{-1}$.

Furthermore, the change in enthalpy decreased more than the change in the entropic term. A comparison of the thermodynamic parameters obtained from Type I and II titrations revealed that there were no significant differences with a change in the mixing sequence. Thus, both mixing protocols were approximately equal in terms of energy release during the first process.

For the second process in Type II c) and Type I d), the Gibbs free energy was also negative and larger than that observed in the first process, indicating that the formation of a coacervate phase was more favorable than the formation of polyelectrolyte complexes. For Type II experiments, $\Delta G_b^B$ varied between $-30.3$ and $-29.5$ kJ mol$^{-1}$, and for Type I experiments, $\Delta G_b^B$ varied between $-30.6$ and $28.3$ kJ mol$^{-1}$. In contrast to what occurred in the first step, in this case, there was more significant change in the parameters according to the mixing protocol. The most significant change was the change in enthalpy because the formation of the coacervate phase for Type II titration was endothermic, whereas for Type I, it was exothermic. Nevertheless, both mixing protocols provided a favorable process in terms of the contribution of the entropic term to the Gibbs free energy, which was enough to make the process possible.

## Ellipsometry analysis

Spectroscopic ellipsometry is a technique that measures the change in the state of light polarization when interacting with a surface. It is suitable for obtaining optical properties, such as the refractive index (n) and extinction coefficient (k), and for accessing the thickness of single- or multi-layered thin films by using mathematical models.[40-42] The interest in using this technique has been increasing owing to its high precision and fast, non-invasive, and non-destructive measurements.[41]

In this study, polyelectrolyte layers were grown on the surfaces of silicon substrates that were assumed to be negatively charged, according to the literature[43] and non-published data by our research group, using layers of polyacrylic acid and the PDADMAC polymer. At pH 10, the first to fifth layers were initially applied to the surface of the silicon using a solution of PDADMAC and sodium alginate. Thicknesses ranged from 0.2 to 2 nm according to ellipsometry parameters $\psi$ and $\Delta$ adjusted by models using the Cauchy, Lorentz, and Cauchy dispersion laws with Urbach tail. Because these layers were so thin, the n values and layer thickness could not be independently determined. Then, n was fixed as 1.46 ($\lambda$ = 632.8 nm) to determine the thickness of the thin layers achieved at pH 10.[41,34]

Additionally, from the ellipsometric data, parameter k from the formed films was defined as a parameter that determines how much a material absorbs electromagnetic radiation of a given $\lambda$.[41] All experiments showed extinction coefficients close to zero, except those determined for 15 or 20 layers in the solutions at pH 3 and 6, respectively. Thus, most films were transparent and did not absorb electromagnetic radiation, justifying the obtained extinction coefficients being close to zero (see Supporting Information S4 for the n and k values for the films formed with 1 to 5 layers at pH 10 and 5–20 layers at pH 3, 6, and 10. S5 for a comparison of the cos $\Delta$ curves for 5 and 20 layers deposited at pH 3 and 10; S6, S7, and S8 for tan $\Psi$ and cos $\Delta$ obtained at pH 10, 3, and 6, respectively). The dispersion laws of Cauchy, Tauc–Lorentz, Gauss, and Cauchy with the Urbach tail and EMA phase were employed to adjust the data obtained for the layers prepared at pH 3 for 5 to 15 layers and at pH 10 for 5 to 20 layers to determine the thickness of the formed layers. Only the Cauchy, Adachi DHO, and Tanguy dispersion laws and diffusion phase were applied for the 20-layer film obtained at pH 3, which exhibited the greatest thickness. The film thicknesses and the coefficient of determination, $R^2$, obtained from each adjustment are shown in Table 2.

**Table 2.** *Thicknesses of layers made of PDADMAC and sodium alginate at different pH values as determined by ellipsometry.*

| pH | Number of layers | Thickness Nm | $R^2$ |
|----|------------------|--------------|-------|
| 3 | 5 | 2.50 | 0.99 |
| | 10 | 42.0 | 0.97 |
| | 15 | 80.5 | 0.91 |
| | 20 | 337 | 0.95 |
| 6 | 5 | 3.10 | 0.98 |
| | 10 | 43.4 | 0.98 |
| | 15 | 61.8 | 0.91 |
| | 20 | 123 | 0.90 |
| 10 | 5 | 2.10 | 0.98 |
| | 10 | 13.4 | 0.98 |
| | 15 | 24.0 | 0.95 |
| | 20 | 44.5 | 0.90 |

The thickness of the layers varied with pH. PDADMAC is a strong polyelectrolyte. It displayed no change in the degree of dissociation as a function of pH and the $-NR_2(CH_3)_2{}^+$ groups were fully positively charged. However, alginate is a weak polyelectrolyte; thus, when the pH of the solution is below its pKa value ($\sim$ 3.2 to 4), there is a greater amount of carboxylic acid groups, i.e., -COOH, for high pH values, and deprotonation allows the formation of ionized groups (-$COO^-$).[44] According to Elzbieciak–Wodka and co-workers[19] and Rydzek and co-workers[18], the behavior of the layer growth at basic pH (pH 10) is associated with a high repulsion between the charged groups ($COO^-$) in the alginate chains. This leads to more elongated chains with thinner layers displaying a typical linear growth. Consequently, this type of growth is usually associated with more structured film formation with poor hydration based on strong interactions.[19,18] At pH 6, the number of deprotonated $COO^-$ groups decreased, and the repulsion was reduced, making the polymeric chains less elongated and the layers slightly thicker. As a result, the growth regime changed to linear behavior. Finally, at pH 3 (acidic pH), there is almost no charge repulsion between the segments and polymer chains that are highly coiled, making the film even thicker.[44] Additionally, the factor that most strongly characterizes this growth regime as exponential is the presence of diffusion between the chains. This type of growth forms less structured films with weaker interactions that are more hydrated and have a higher permeability tendency.[19,18]

According to Decher and Schlenoff[45] based on studies by Yoo and co-workers[46] and Durstock and Rubner[47], the pH can affect layer formation when PAA/PAH are used as weak polyelectrolytes. The non-ionization of -COOH groups by using dipping solutions at low pH lead to the formation of non-stoichiometric pairing of the repeated units in the polymers chain. In this case, our result for a multilayer or surface with excess sodium alginate and low density of ionic crosslinks produces thicker layers. For the dipping solution with a high pH, sodium alginate had a completely ionized chain that resulted in a structure with more ionic crosslinks and a more equilibrated ratio of polymers, which produced thinner layers. Another factor that may explain the exponential growth of the film is the formation of coacervate complexes, a type of gel formed owing to the sequential deposition of layers.[48] According to Elbert and co-workers[49], who studied thin polymer films formed on models of tissue surfaces using polyelectrolyte multilayer techniques, during the process of constructing multilayers of polylysine and alginate at a determined concentration and pH, a gel formed at the interface of the layers when the growth regime was exponential. This gel consisted of coacervate complexes. When the conditions (pH and concentration) changed, the growth regime changed, and gel formation was not observed.[49]

Another factor that influences film thickness is the ionic strength. According to Maza and co-workers[50], who studied the poly(4-styrenesulfonic acid-co-maleic acid) (PSS-MA)/PDADMAC assembly, there is a dependence between film growth at different pH values and the presence of salt in the polyelectrolyte solution. This is associated with the ionic strength; if the system under analysis contains salt in the polyelectrolyte solution, then a charge screening effect in the polyelectrolyte chain segments is observed. Thus, the charge density decreases, forming more coiled chains. However, in systems that are salt-free, as in our case, the ionic strength remains the same, and there is no charge screening effect. Thus, the formed layers are thinner and stratified. For the type of system that is salt-free, Maza and co-workers[50] observed that the thickness also decreased as the pH increased, which is in line with our studies. In our study the increase in layers as a function of pH did not cause considerable variations in the optical parameters of the formed films, except for in the 15-layer film obtained at pH 3, which showed different aspects with some coloration (see Supporting Information S9). As already mentioned, the k parameter is related to the absorption of light of the film[41]; thus, it is expected to vary in relation to other films that are transparent. The coloration of this film can be explained by the interference effect. When light is directed to the sample surface, it interacts with it and is reflected from the film surface and the film/substrate interface. Thus, there is an overlap between the reflected beams. If such an overlap of the reflected beams is in phase, then constructive interference occurs, but if they are out of phase, destructive interference occurs.[41] As films increase in thickness, the beam overlap changes, and this interference can be altered, thus affecting the wavelength at which the films absorb.

## Atomic force microscopy analysis

Atomic force microscopy measurements were performed to evaluate the film deposition and compare it with the ellipsometric results. Based on the obtained scans, we obtained a topographic image from which mainly data on the thickness of the films were extracted. From the cropped images, especially 3D images (see Supporting Information S10), the region of the film near the cropping was much thicker than the rest of the film. This was observed because, at the cut (scratch), a drag of the film occurs, which results in material accumulation in the edges of the scratched area. Thickness measurements were thus performed that excluded the edges of the film to avoid drag influence. Because the films showed considerable heterogeneity in film thickness, especially the thicker ones, thickness measurements were performed in the medium height regions of the film. From the AFM topography images and thickness measurements, the deposition of the films was non-uniform, presenting forms with a relief that was initially soft and became more protuberant and heterogeneous as the number of layers increased, which likely also resulted in increased roughness. This behavior is probably associated with the stability of the base layers that more closely adhered to the substrate and were consequently more stratified. However, this characteristic was lost as the number of layers increased. Some factors may also influence the film homogeneity, such as the type of polymer and its polydispersity, pH, and presence of salt.[51]

## Relation between pH and growth regime

To evaluate the growth behavior of the layers according to the increased deposition of layers, a plot of thickness versus the number of deposited layers was obtained for each pH used during the deposition procedure. Figure 7 shows a comparison of the thickness results obtained using AFM and ellipsometry. The results were in good agreement, except for the film with 20 layers obtained at pH 3. This may have been related to the heterogeneity of the surface, as observed by AFM.

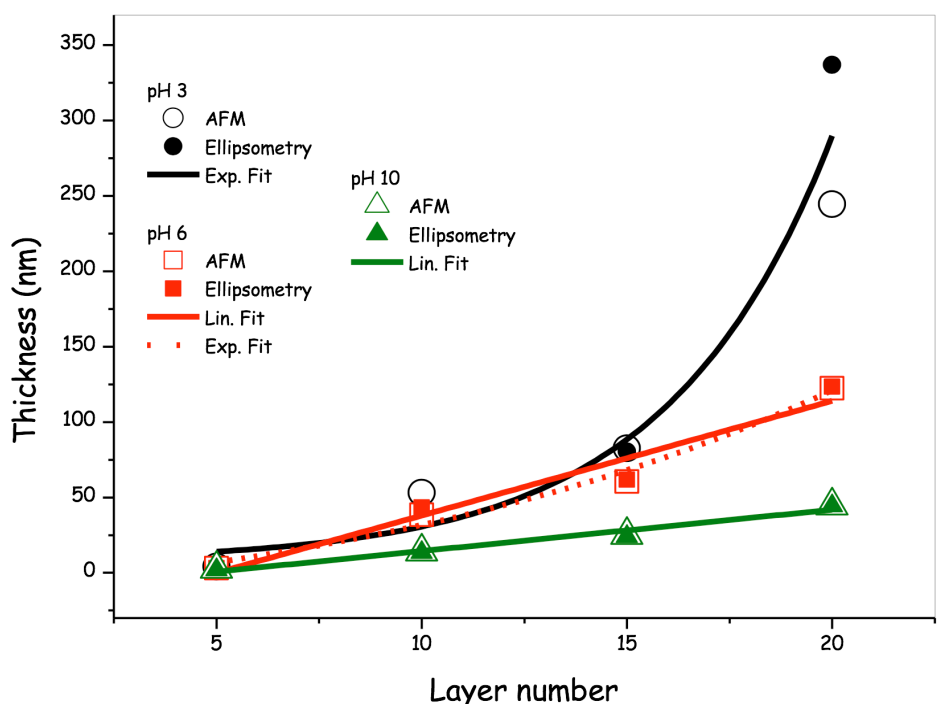

***Figure 7.*** *Growth regime of PDADMAC and sodium alginate films at pH 3 ($R^2_{AFM}$ = 0.92 and $R^2_{El.}$ = 0.98), pH 6 ($R^2_{AFM, lin.}$ = 0.97 and $R^2_{El., lin.}$ = 0.94; $R^2_{AFM, exp.}$ = 0.98 and $R^2_{El., exp.}$ = 0.92), and pH 10 ($R^2_{AFM}$ = 0.98 and $R^2_{El.}$ = 0.96). $R^2$, El., lin., and exp. denote the coefficient of determination, ellipsometry, linear, and exponential, respectively.*

As shown in Figure 7, the layers showed different growths according to the pH used during the layer-by-layer deposition, and the rate of growth increased as the solution pH decreased. The layers obtained at pH 3 showed exponential growth behavior, whereas at pH 10, the layers showed single linear growth behavior. Layers grown at pH 6 showed an intermediate growth regime between linear and exponential. Laugel and co-workers[14] associated the grown films with the thermodynamic parameters related to the association between the polyelectrolytes. Endothermic complexation was associated with exponential growth, and the strong exothermic process was associated with linear growth, whereas the weak exothermic process was associated with weak exponential growth. Alonso and co-workers[39] showed that the dependence of the pH on the layer formation of PAA and PDADMAC was also associated with the thermodynamic parameters of the association. At pH values of 3 and 6, the association process was exothermic, and at pH 10, the process was endothermic. For the endothermic process at pH 10, linear growth was observed, and supralinear growth behavior was observed in acidic pH.

In our study of sodium alginate and PDADMAC, independent of pH association, the process was endothermic, and dual behavior was observed. At low pH, exponential growth was observed, and at high pH, linear growth was observed. The growth behavior was different even when the thermodynamics of the process was similar and this behavior has not been observed yet. However, Richert and co-workers[52] showed that by decreasing the electrostatic interaction by adding salt, films of hyaluronic acid and chitosan at low salt concentration showed a linear regime of growth.

Increasing the salt concentration and decreasing the electrostatic interaction between the polyelectrolytes resulted in exponential grown, according to their report.

**Interferometric confocal microscopy**

Figure 8 shows the behavior of the roughness of films with 5, 10, 15, and 20 layers grown at different pH values for the dipping solutions. Generally, the highest values of roughness were observed for layers grown at pH 3, followed by those grown at pH 6 and 10. Additionally, there was an increase of roughness with an increase in the number of layers deposited. For pH 3 for all added layers, there was a significant increase in the roughness. For others pH this effect was less intense.

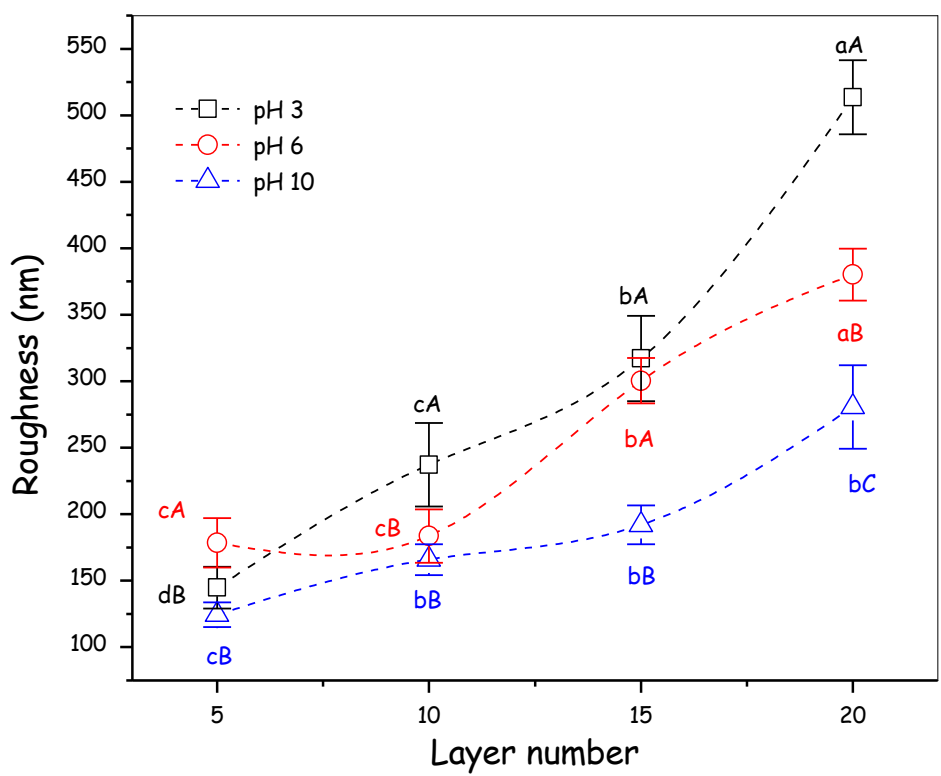

**Figure 8.** *Roughness of films according to interferometric confocal microscopy of the PDADMAC and sodium alginate films produced at pH 3, 6, and 10 with 5, 10, and 20 layers. Equal lower case letters show that for the same pH, the results did not differ according to the Tukey test at a 5% significance level. Equal capital letters show that on the same layer, the results did not differ according to the Tukey test at a 5% significance level.*

Shiratori and Rubner[53] studied pH-dependent thickness behavior of sequentially adsorbed layers and observed that the surface roughness has a considerable influence on the multilayer film thickness. They observed that an increase in thickness is accompanied by an increase in roughness. As previously mentioned, the formation of thicker layers is related to the presence of coiled polymeric chains. This coiled conformation is dominated by a large amount of loop and tail segments, and this type of dipping provides an increase in surface roughness.[53,44] At pH 3, the alginate polymeric chains were more coiled because the repulsion between the chains was weak; consequently, there was a large amount of segments in the loop and tail, which then provided the high roughness shown in these films. At pH 6, the alginate chains lost part of the coiled characteristic owing to the increase in the charge density, which increased the repulsion; thus, the number of segments in the loop and tail decreased, also causing the surface roughness to decrease. Finally, at pH 10, the chains were fully elongated due to the high repulsion so that the segments had a flat shape, which reduced the surface roughness.[53,43] Figure 9 shows the 3D topographical images of the 20-layered film for layers grown at pH 10, 6, and 3, respectively, in a, b, and c. Figures a and b show films from dipping solutions at pH 10 and 6, respectively, that presented a homogeneous surface, whereas the sample from pH 3 presented a more irregular surface. However, some holes were observed in both samples, as shown in black color (see Supporting Information S11 for the topographic images of the films by interferometric confocal microscopy for the different layers).

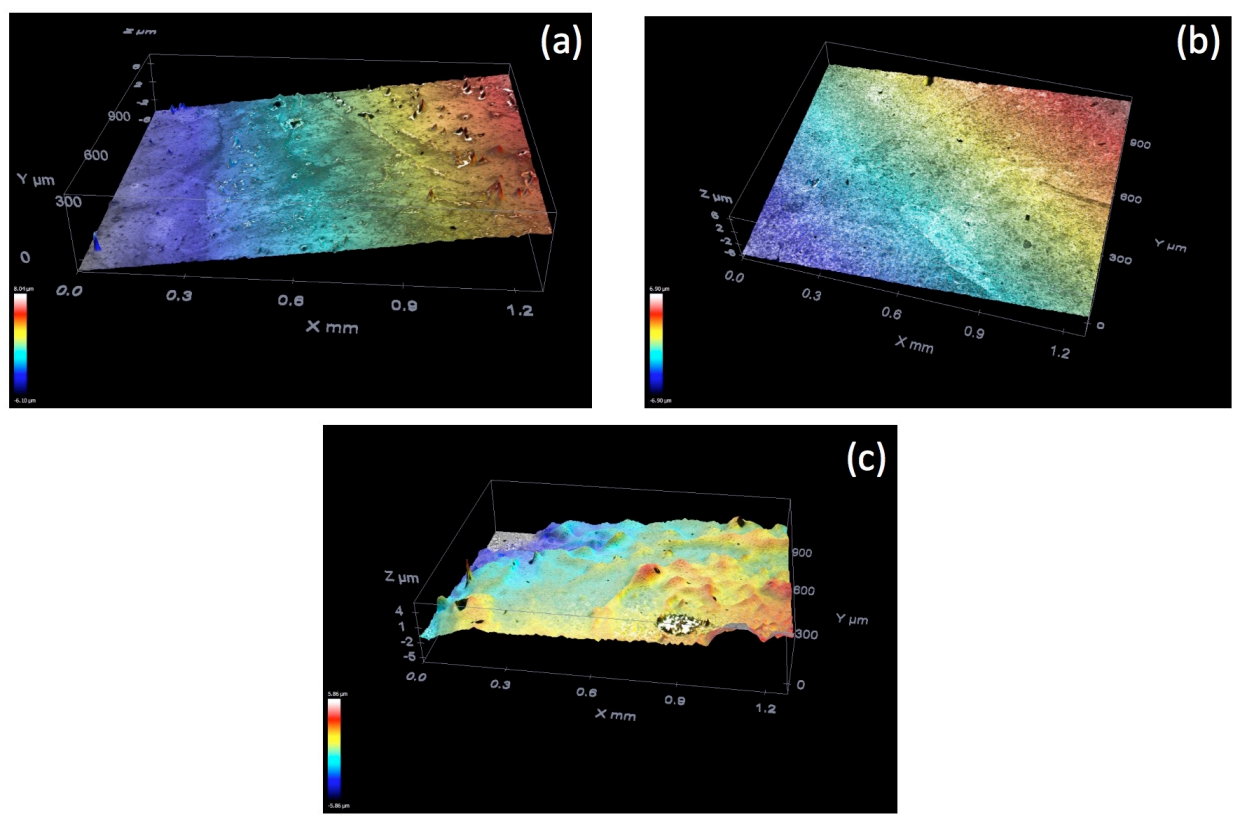

**Figure 9.** *3D representative topographical image of films from the 20th layer by interferometric confocal microscopy of PDADMAC and sodium alginate films deposited at (a) pH 10, (b) pH 6, and (c) pH 3.*

By analyzing the roughness values and comparing them with those of the topographic images, we observed that the data agree. At pH 3, the roughness values shown are higher, which can be confirmed by the more irregular surface, as shown in the topographic image in Figure 9c. At pH 6 and 10, the lower roughness values than those observed at pH 3 agree with the topographic im-ages in Figures 9b and 9a, respectively, which show more homogeneous surfaces. Thermodynamic data from the association of polymers in solution suggest that for the layers formed by the addition of sodium alginate to PDADMAC, the thermodynamic parameters of the first process, or process A, are related to the polymer–polymer interaction.

There was an increase in $\Delta H$ with an increase in pH with a corresponding increase in the entropy of this association. Thus, more water or ions were likely released during the formation of the polymer–polymer aggregate. In the formation of the layer, higher entropy values justified the formation of denser and less hydrated layers at a higher pH. This may justify the formation of surfaces with less roughness for complexation made with dipping solutions at more basic pH.

However, when PDADMAC was added to sodium alginate at pH 4 or 5, the DH values were higher than those for the respective experiments when alginate was added to PDADMAC. For pH 7 or 10 in which the alginate was more deprotonated, the values of $\Delta H$ and $\Delta S$ were inverted and were higher for the addition of alginate to PDADMAC. Therefore, there may be a change in the surface characteristics when the layer is grown with the addition of sodium alginate or the addition of PDADMAC. As previously suggested by Liu and co-workers[36], the coacervation process is associated with an energy cost of achieving condensation of aggregates in drops of neutral coacervates. According to the thermodynamic data, the formation of coacervates during the addition of alginate in PDADMAC was favored entropically by the release of species in the aggregates into the water. However, when coacervates were formed by adding PDADMAC in sodium alginate, the process was energetically favored because a negative DH was observed. More specific studies can be performed in the future to strengthen this hypothesis.

## CONCLUSION

In this study, we investigated the electrostatic complexation of cationic and anionic polyelectrolytes, PDADMAC and sodium alginate, respectively, by combining data from dynamic light scattering and isothermal titration calorimetry on the co-assembly with those from studies of polymer solutions and the formation of layers whose thicknesses were characterized by ellipsometry and atomic force microscopy. Regardless of the mixing sequence of the polymer solution, the association process occurred in two steps: 1) a process of aggregation that formed polyelectrolyte complexes, with smaller sizes and 2) a process with phase separation and formation of particles that were coacervates or precipitates. The ITC results revealed that the sequence of addition of the polyelectrolytes did not interfere thermally in the first-step process of complexation, which was endothermic in both directions in the entire range of pH studied. However, for the first-step process, the $\Delta H_b^A$ became less endothermic as the pH decreased even though both processes were entropically driven, i.e., favorable, which may be related to the large mobilization of water molecules in the structure of the polymers that favor complexation.For the second-step process, the order of addition of the polyelectrolytes interfered with the signal of $\Delta H_b^B$. Type II titration was endothermic, and Type I was exothermic. For the addition of sodium alginate to PDADMAC, the Gibbs free energy was low for the neutral solution for the first-step process and for the acid solution for the second-step process at pH 7 and 4, respectively.

For the addition of PDADMAC to sodium alginate, similar behavior was observed; the Gibbs free energy was lower for a neutral solution for the first-step process, and for the acid solution for the second-step process at pH 7 and 5, respectively. Thus, a neutral medium favors the formation of a polyelectrolyte complex between sodium alginate and PDADMAC for both mixing protocols. However, coacervation was more favorable in an acid medium between 4 and 5. The combination of DLS and ITC data was useful for understanding the pH influence on the mechanism of association of the polyelectrolytes.The film thicknesses obtained by ellipsometry and AFM were in good agreement. The results obtained showed a pH influence on both thermodynamic aspects of the association and characteristics of the obtained layers. At basic pH, a linear growth regime was obtained, and as the pH became more acidic, the growth of the layers became exponential. These particularities help us to understand the structures generated from the association of the polymers employed and can thus better direct the applicability of the films and the conditions necessary to reproduce them. The analyses from interferometric confocal microscopy showed the influence of pH and the number of layers deposited in relation to the roughness of the films. For the growth of multilayers of PDADMAC and sodium alginate, the roughness increased as the number of deposited layers increased; thus, the thicker films exhibited greater roughness. Additionally, the roughness increased as the pH decreased, i.e., the films were rougher at acidic pH and less rough at basic pH.

## SUPPORTING INFORMATION

Viscosimetric determination of molar mass of sodium alginate; Procedure for cleaning the silicon substrate; Evolution of the pH along the titration; Refractive index (n) and extinction coefficient (k) for films with 1 to 5 at pH 10 and 5, 10, 15 and 20 layers from dipping solutions at pH values of 3, 6 and 10; Curves of cos Δ for 5 and 20 layers deposited at pH 3 and 10; Curves of tan (Ψ) and cos (Δ) for samples at pH 10; Curves of tan (Ψ) and cos (Δ) for samples at pH 3; Curves of tan Ψ and cos Δ for samples at pH 6; Pictures of the samples made at different pH values of dipping solutions; Topographic images of the films by AFM; Topographic images of the films by interferometric confocal microscopy and References.

## AUTHOR CONTRIBUTIONS

*Authors Macedo, B. S. and Almeida, T. contributed equally to the experimental part of this study.

## ACKNOWLEDGMENT

We thank the staff members of the Multiuser Laboratory of Material Characterization/Brazil (www.uff.br/lamate) for training and assistance pertaining to the VP-ITC microcalorimetry (GE) results included in this publication, the Microstructural Characterization Laboratory/Brazil for the AFM measurements, and E. E. Garcia-Rojas and W. Loh for the use of the zetasizer. Almeida, T. and Macedo, B. S. thank CAPES/Brazil for the Master Degree scholarship. Netto, A. D. P. thanks CNPq/Brazil for the individual research grant (Process 312288/2017-4). L. Vitorazi thanks FAPERJ/Brazil for funding (Process E-26/202.724/2019 and E-26/010.000982/2019). We

also thank FINEP for the acquisition of the ellipsometer (SEMILAB®, model SOPRA GES 5S) and the interferometric confocal microscope (LEICA®, model DCM3D).

# LANGMUIR
The ACS journal of fundamental interface science

## Graphical abstract

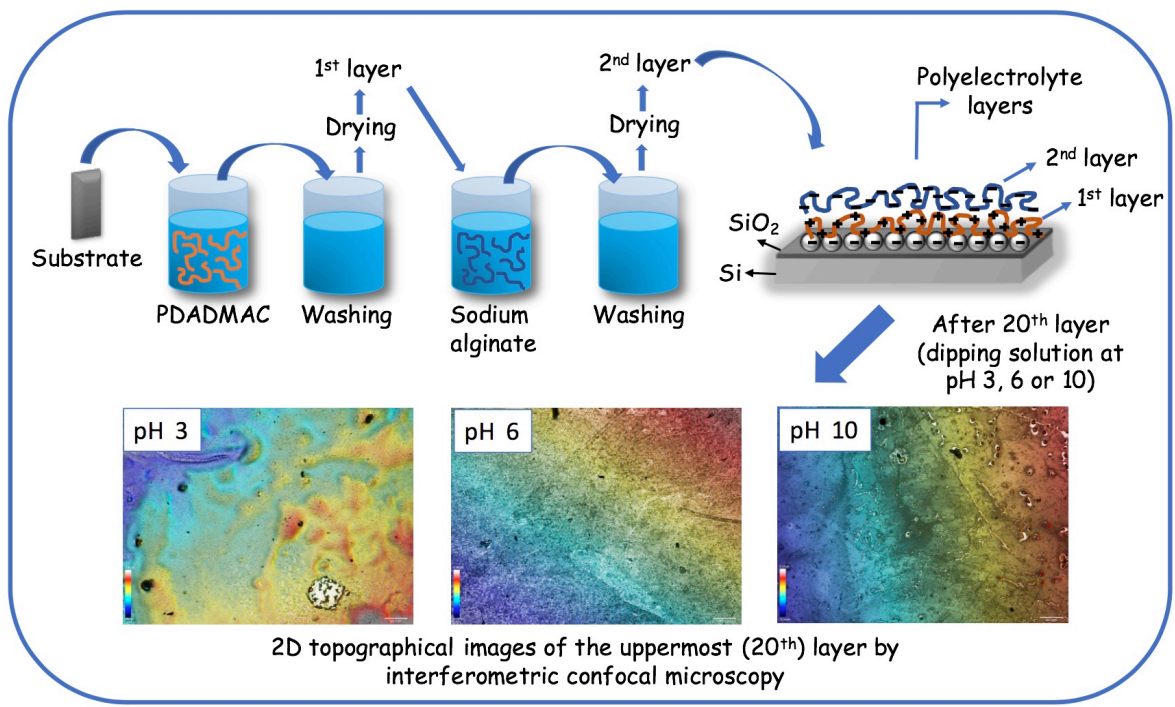

2D topographical images of the uppermost (20th) layer by interferometric confocal microscopy